\documentstyle[11pt,newpasp,twoside,epsf]{article}
\begin{document}
\title{Age dating old globular clusters in early-type galaxies}
 \author{Markus Kissler-Patig}
\affil{European Southern Observatory, Karl-Schwarzschild-Str.~2, D-85748
Garching, Germany}

\begin{abstract}
Various methods for age dating globular clusters in ellipticals are
presented. We first present {\it spectroscopy} of individual globular clusters
(feasible with the advent of the 10m-class telescopes), and the measurement of
Balmer line indices. Second, we discuss the {\it photometry} of globular
cluster sub-populations and the mean age determination by comparison
with population synthesis models.
The first method is time consuming but precise once spectra with high enough 
signal to noise are obtained. One caveat, however, is the definition of the
Balmer line indices that often include metal features and are themselves not
independent of metallicity.
The second method requires the measurement of two photometric quantities
that depend differently from age and metallicity. The combination of
both allows to break the age-metallicity degeneracy present in
broad-band colors. Near-infrared colors can strongly complement the
optical studies in this respect.

Overall it is, however, true that the older a star cluster, the harder it
is to determine its precise age. Population synthesis models also need
to be further improved, and the models compared here show significant 
differences.
\end{abstract}


\section{Introduction}

\begin{figure}[b]
\plotfiddle{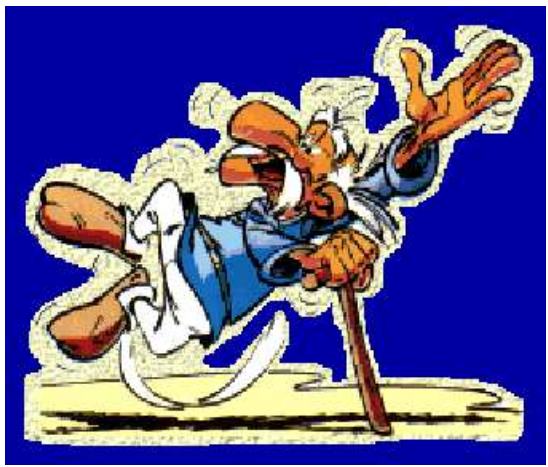}{5cm}{0}{60}{60}{-175}{-135}
\caption{The older an object, the harder it gets to guess its age}
\end{figure}

Globular clusters can be old. Very old. Sometimes older than the
universe. This is then referred to as a cosmological paradox, and it
indicates that you are using wrong values for your favorite cosmological 
parameters. But how old are they really and why do we care? 

Age dating old stellar populations has never been trivial, but
crucial for our understanding of the universe. For example, the ages of
Galactic globular clusters are still the reference for the age of the
universe. This aspect of the problem was recently discussed in several reviews
(e.g.~Salaris, Degl'innocenti, \& Weiss 1997; Sarajedini, Chaboyer \&
Demarque 1999 and references therein, see also Weiss et al.~in
these proceedings).

Here, we would like to present the current methods (from a purely observational
point of view) to determine ages of star clusters in galaxies
beyond the Local group. In other words: how to determine ages from the
integrated light of star clusters. These ages are used to date the major
epochs of star formation in the host galaxies, and thus to constrain the
formation history of these same galaxies.

The next section presents age determinations from spectroscopy, with its
advantages and problems. In section 3, we present photometric methods
to derive relative ages of populations, and their caveats. A few
concluding remarks are given in Section 4. 

\section{Spectroscopy}

Spectroscopy of the integrated light of globular clusters
allows the age determinations of {\it individual} clusters by measuring
various absorption line indices in their spectra. The faint magnitudes of the
objects prohibit high-resolution spectroscopy. Furthermore, one would
like to compare the results with older/other ones, i.e.~use a `standard'
system (e.g.~the Lick indices) which were measured on low-resolution
spectra (6\AA\ to 9\AA\ resolution) in the wavelength range typically
ranging from 3800\AA\ to 6500\AA, sometimes including the Ca triplet around
8500\AA.

\subsection{The real life}

In order to spectroscopy globular clusters, one needs to identify them
in the first place. Therefore, each spectroscopic survey must be
preceeded by a photometric one. All studies are still contaminated to some
extend by foreground stars (mainly M stars) and compact background
galaxies at low redshift.  Out to a distance of $\sim$ 30--40 Mpc
the globular clusters can be (barely) resolved with WFPC2 on HST, which
is currently the best method to prevent contamination by foreground
stars, when associated with a color selection. However, HST photometry 
is usually
available for a small field only. Multi-color, wide-field photometry, especially
over a large color baseline, can select out background galaxies
efficiently. Current studies that use both these methods to select their
globular cluster candidates have typical contamination rates of less
than 10\%--20\%. Further, the advent of 10m-class telescope allow the
spectroscopy of fainter objects, i.e.~the choice of ``secure'' globular
candidates within the spectroscopic field increased significantly,
making the modern studies even more efficient.

Still, the method is time consuming: typical exposure times to get
a useful signal-to-noise to derive ages are around 2--3h, even with 10m-class 
telescopes. Also, the current multiplexity of FORS~1\&2 (VLT) and LRIS
(Keck) is low with around 20--30 candidates per set-up. Instruments such
as VIMOS (VLT) and DEIMOS (Keck) with a multiplexity of around 100--150
objects/set-up will significantly improve the efficiencies of such studies. 

\begin{figure}
\plotfiddle{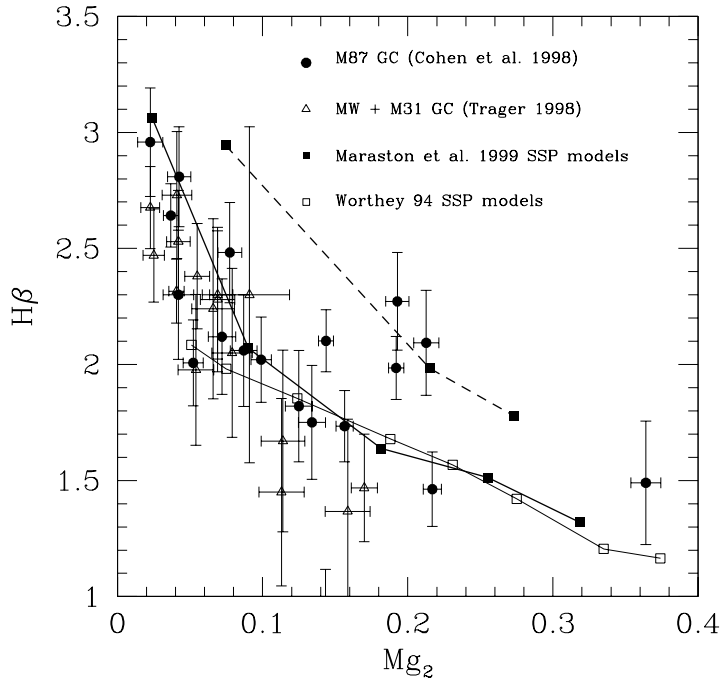}{8cm}{0}{50}{50}{-160}{-100}
\caption{H$\beta$ plotted against Mg$_2$ for a number of Milky Way, M~31
and M~87 globular clusters. Overplotted are models of Maraston (1998)
for 15 and 7 Gyr, and [Fe/H] ranging from -2.25 to 0.35, as well as
models from Worthey (1994) for 17 Gyr and metallicities ranging from
-2.0 to 0.5.}
\end{figure}

Once spectra are obtained, absorption line indices are measured. In
order to compare these with existing population synthesis models, one
uses a standard system of absorption line indices. This
standard system was implicitly adopted to be the Lick system 
(see Trager 1998 for the latest update), in which a large number
of early-type galaxies (and globular clusters) were originally measured.
The index measurement of Balmer lines are finally compared to population 
synthesis models and the ages are derived.

\subsection{Problems and accuracy}

The last sentence is overly optimistic and hides many problems, some of
which are illustrated in Fig.~2. First, H$\beta$ is the index mainly used 
to derive ages.
This line index is not a pure age indicator but also sensitive to metallicity 
(half as much as it is sensitive to age, see Worthey 1994). This is 
illustrated in Fig.~2
by the Milky Way data which are roughly coeval and should lie on a
horizontal line if H$\beta$ was a perfect age indicator. In contrast,
they describe an almost vertical feature at low metallicities. This is
the reason why, in order to determine ages, H$\beta$ is generally
plotted against a very metal-sensitive feature (here Mg$_2$) to
understand and take into account the metallicity contribution to
H$\beta$. Ideally, an age sensitive
feature would mainly probe the turn-over in the Herzsprung-Russel
diagram. Unfortunately, H$\beta$ is very sensitive to blue horizontal
branches too. This is especially a problem in metal-poor population
(typically exhibiting an extended horizontal branch), but could also
falsify result in metal-rich population hosting blue horizontal branch
stars. H$\beta$ is therefore sensitive to some extend to the second parameter 
effect. We note here that Worthey (1994) did not include any blue
horizontal branches in his models. Finally, H$\beta$ can be
contaminated by line emission, although this is less of a concern in the case
of old globular clusters but rather for galaxies with some recent star
formation. And last it should be noticed that the ``narrow'' definition
of H$\beta$ is not well suited to measure the broad Balmer lines of
young population.

The second big practical problem is the accuracy with which H$\beta$ is 
currently measured. Figure 2 illustrates that current typical errors for 
distant globular clusters are of the order of a factor 2 in age. Spectra
with higher signal-to-noise must be obtained. Previous studies were
either limited by the telescope size, or by the blue response of the
used CCDs, or simply optimized for metallicity measurements between
5000\AA\ and 6000\AA . Future studies must concentrate on getting enough
signal-to-noise below 5000\AA\ which is feasible with the advent of
10m-class telescopes. Also, if the goal is to get a relative age difference
between sub-populations of (rather than individual) globular clusters,
the measurement of enough representative clusters
currently allows a determination to within 2--3 Gyr.

The third problem, already mentioned, is the models to which the data
are compared. Current population synthesis models do not agree on the absolute
ages derived from line indices. Worse, they also influence the relative
ages: i.e.~the spacing between the different isochrones in Fig.~2 varies from 
model to model. 

The three points above should make clear that in order to get relative ages
between individual clusters or cluster sub-populations one needs to work
on the following points. Other age sensitive features should be used
(e.g.~higher-order Balmer lines) with a new definition of the indices, 
perhaps at higher resolution, in order to avoid metal lines in the
definition of the bands. These features need to be measured with a higher
signal-to-noise than is currently done. And finally, the population
synthesis models need to be brought in agreement with each other.

\section{Photometry}

Spectroscopy being extremely time consuming, ways of measuring ages for
old globular clusters from photometry were explored.

\subsection{The basic idea}

Broad-band colors suffer from the well known age-metallicity degeneracy.
That is: younger ages can be compensates by higher metallicities. The
goal is therefore to find two photometric quantities that depend
differently on age and on metallicity in order to break the degeneracy.
Broad-band colors at old ages are all much more metallicity sensitive
than they are age sensitive, and any combination (from U$-$V to V$-$K)
plotted against each other is unable to separate ages.
 
A solution to this problem was to combine a color with a magnitude.
Colors are, as mentioned above, metallicity sensitive, while magnitudes
are rather age sensitive. Plotting the one against the other breaks the
age-metallicity degeneracy and allows the determination of relative ages.

\begin{figure}
\plottwo{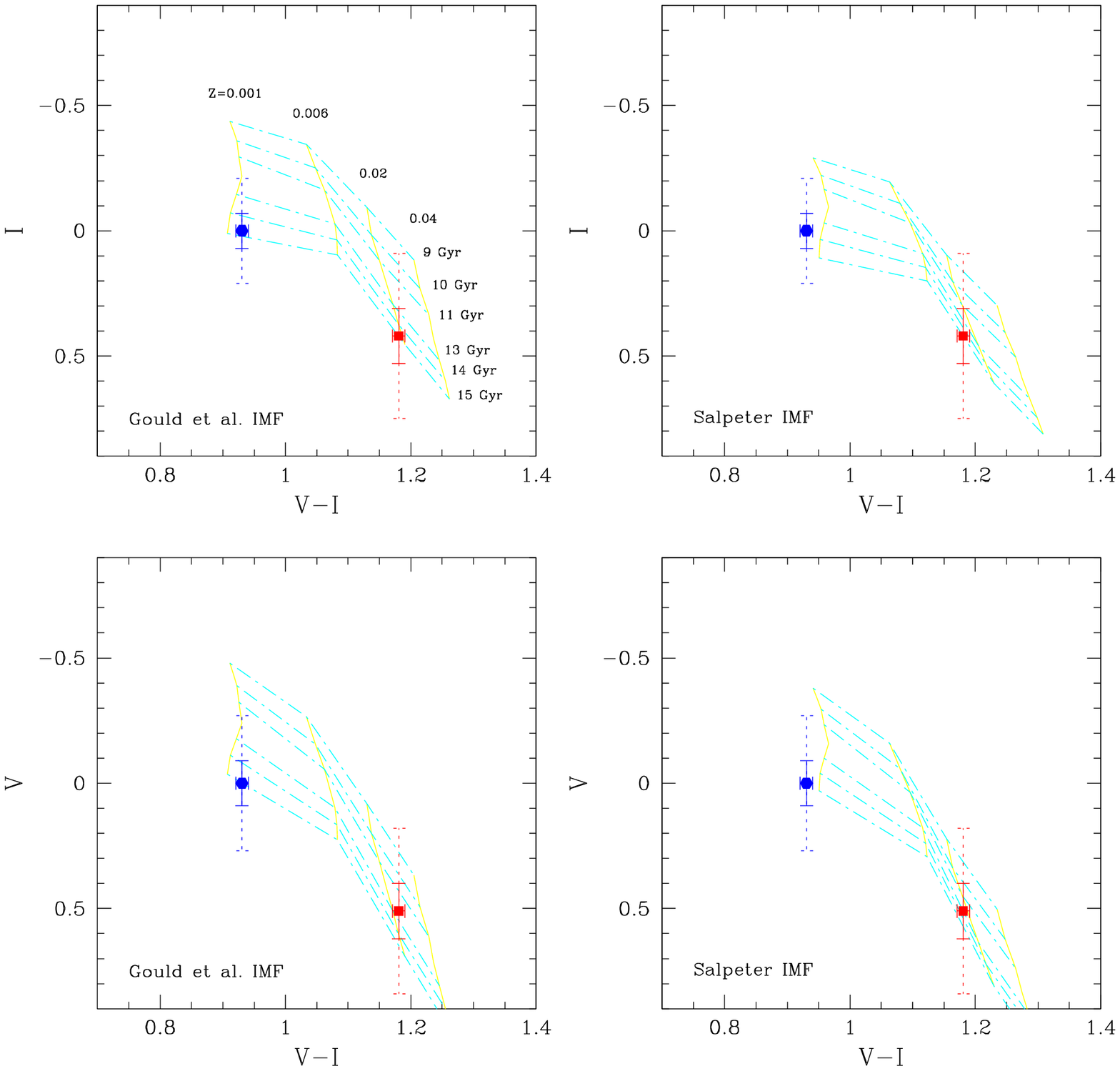}{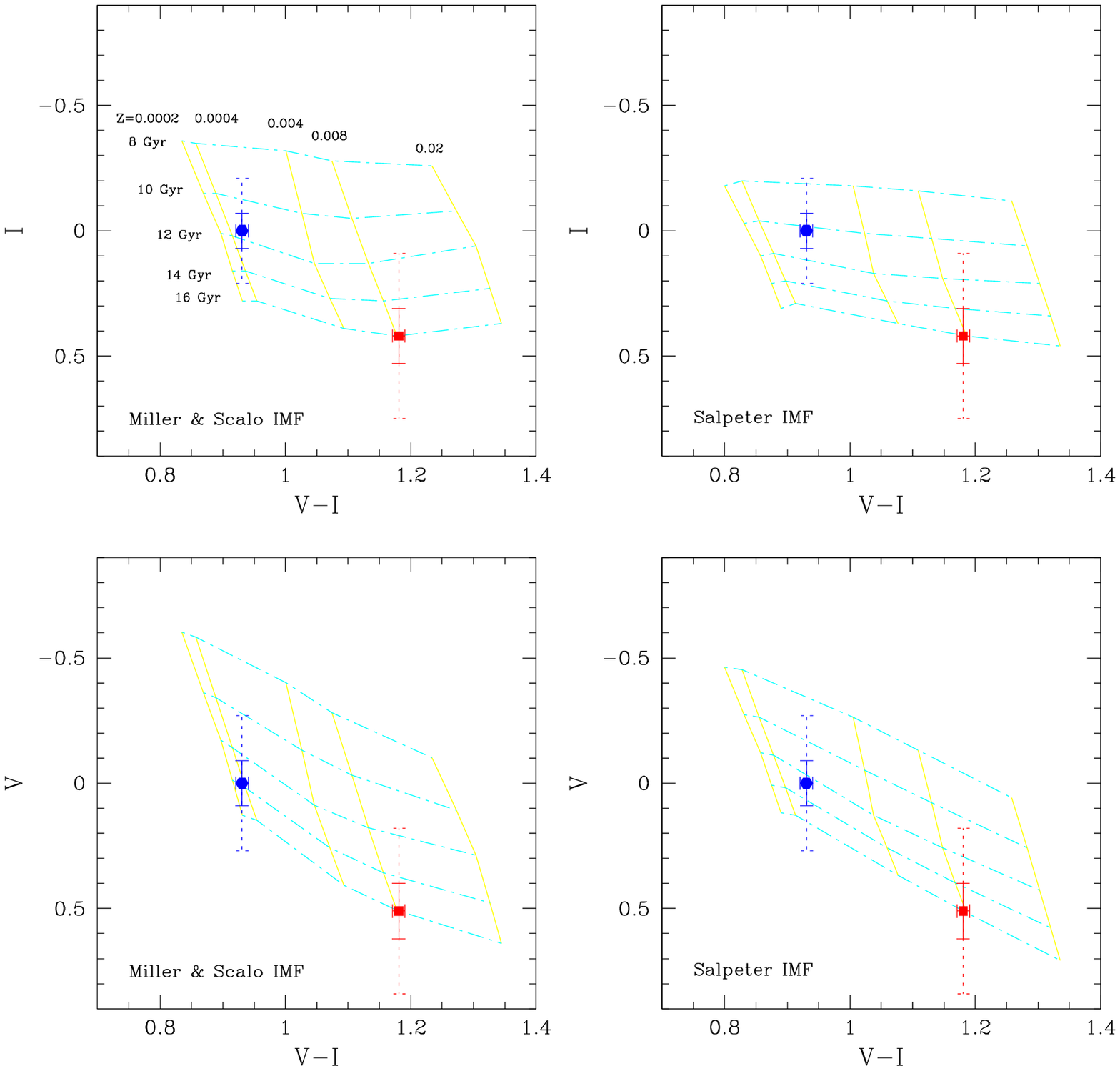}
\caption{Comparison of the Models of Maraston (1998; left) and Worthey
(1994; right). The blue and red points are the mean color and ``mean'' 
magnitude for the metal-poor and metal-rich globular cluster populations in 
NGC 4472. The plot is take from Puzia et al.~(1999)}
\end{figure}

\subsection{Some results and caveats}

As for spectroscopy, this method sounds easier than it is in reality.
First, the individual magnitudes depend of course primarily on the mass of 
the globular clusters. 
The masses being unknown or dependent on the exact distance,
mass-to-light ratio etc... this method cannot be applied to individual
clusters. However, one can determine a mean color and a ``mean''
magnitude for a globular cluster population. The mean color is
just the peak of the color distribution of the cluster population.
The ``mean'' magnitude is taken to be the turn-over magnitude of
the luminosity function, corresponding to the characteristic mass of the
cluster population (see Kissler-Patig; Miller; Fritze-von Alvensleben;
McLaughlin, all in these proceedings). 

The exact value of this characteristic mass is unknown to within
10\%--20\%, and usually the distance to the objects is uncertain by a
similar amount, so that {\it absolute} ages cannot be derived. The
method remains, however, useful to compare the ages of different
sub-populations (e.g.~the case of NGC 4472 by
Puzia et al.~1999). One assumption is that both sub-population have the
same characteristic mass implying that color and magnitude depend on
age and metallicity only. This assumption seems supported by current
theory and observations (see above reference) but can also be checked by
comparing the age difference between the sub-populations, derived from
various filter combinations. The results from various bands
will only agree if the quantities are indeed only dependent on age and
metallicity (fully taken into account by the models), and systematic
differences will appear if the characteristic masses of the two compared 
populations differ. 

Figure 3 illustrates the method. It compares the result derived with
respect to two different population synthesis models. As the error bars
illustrate, it is rather
easy to derive a mean color for a sub-population, however the ``mean''
magnitude is much harder to measure and requires a large sample of
clusters with good photometry and knowledge of the finding
incompleteness as a function of cluster magnitude and background
luminosity. In this case, the relative age between the two populations
can be derived to within 1-2 Gyr.
Note, however, that even the age {\it difference} depends on the
population synthesis models used for the comparison. We picked two
extreme cases that illustrate that the uncertainties in the models do
not allow a determination of the relative ages to better than 3-4 Gyr.
In one case the luminosity is, still in the I band, dependent on
metallicity, while in the other case such a dependence exist in the V band
but disappears in the I band (as seen from the almost horizontal
isochrones). The consequence is, that for the second
model (Worthey 1994) the results from the V band indicate that only a
marginal difference in age exists between the metal-poor and metal-rich
clusters, while the results in the I band seem to favor an older
metal-rich population.

Overall, the comparison (taken from Puzia et al.~1999) supports coeval 
metal-poor and metal-rich populations to within the uncertainties (3-4 Gyr).
Similar, perhaps a little less rigorous, studies came to comparable
conclusions. Kissler-Patig et al.~(1997) found the two sub-populations
in NGC 1380 to be coeval with weak evidence that the metal-rich
population might be 3-4 Gyr younger. Kundu et al.~(1998) find a 
metal-rich population younger by 3-6 Gyr than the metal-poor population
in M~87. 

\section{Concluding remarks}

If there is one thing to remember: getting ages of old globular clusters
from integrated properties is tough! But it can be done. And although
the current methods still have limits, there is good hope that in the
near future we will be able to get relative ages of individual clusters to 
within 2-3 Gyr, and relative ages between groups of clusters to within
1--2 Gyr. Differences in the models currently limit the accuracy
with which relative ages can be determined to 3--4 Gyr.
Absolute ages await an agreement between the different stellar
population models (including a better knowledge of stellar evolution),
and is probably to be expected at earliest in a decade.

When individual ages are required, spectroscopy is the only possible
method. However, for relative ages of entire populations, photometry can
be used. The latter will also allow to determine relative differences in
the characteristic mass between cluster sub-population.

\acknowledgments

I would like to thank the organizers and in particular Ariane Lan\c{c}on
for initiating and organizing such a pleasant meeting.
Special thanks to Claudia Maraston for her patience in trying to
convey some of her expert knowledge on population synthesis models to a
dum observer, and for her plot. Thanks also to Thomas Puzia, responsible
for the figures in the photometry section.

\section*{Discussion}

\noindent {\bf B.Miller:} How does Washington photometry compare to
other bands for determining metallicities of old globular clusters?

\noindent {\bf M.Kissler-Patig:} Washington photometry uses the C
(somewhere between U and B) and T$_1$ ($\sim$ R) bands to derive
metallicities. It is more sensitive to metallicity than B$-$I, the most
sensitive combination of Johnson-Cousins bands, but it is less sensitive
than V$-$K. However, it has the advantage over the latter that it can
be obtained with a single instrument.

\noindent {\bf J.C.Mermillod:} What is the population producing the H$\beta$
line, turn-off dwarfs, blue horizontal-branch stars, or blue stragglers?

\noindent {\bf M.Kissler-Patig:} All of the above. H$\beta$ is
located at $\sim$ 4800\AA . Ideally, you would like it to be dominated
by turn-off dwarfs, but especially for metal-poor populations, the
horizontal branch is a significant contributor. I don't think that blue
stragglers contribute significantly unless you have a very large
population of these, but their represent a uncertainty factor also.

\noindent {\bf J.Gallagher:} I am also concerned about the ability to
separate age and metallicity well enough to be able to say with
certainty that the blue and red clusters are coeval. My problem is that
one can get very similar distributions of stars on a theoretical CMD at
different ages -- including luminosities -- which could lead to
increased uncertainties in relative ages of cluster groups.

\noindent {\bf M.Kissler-Patig:} Indeed, the problem you state is a
concern. However, I think that several points save you from making too
large errors. The first one is, that we are ``averaging'' over hundreds
of clusters (i.e.~CMDs) and are not very sensitive to peculiar effects
in the one or the other clusters (such as weird horizontal-branches
etc...). Second we are starting to use a wide range of colors, probing
very different regions of the CMD (e.g.~the K band will hardly be
sensitive the turn-over region or the horizontal-branch morphology,
while the B band will). We should therefore see unexpected difference between
the bands when compared to the models, if the CMDs differ a lot from
the model predictions. This whole business reduces then to model
uncertainties that are present to some extend as mentioned, but do not
disqualify the method as such, nor the results that the populations are
coeval within a few Gyr.

\end{document}